\documentclass[reprint,aps,prl,twocolumn,superscriptaddress,nofootinbib,floatfix,longbibliography]{revtex4-2}

\usepackage[T1]{fontenc}
\usepackage{amsmath,amssymb,bm}
\usepackage{graphicx}
\usepackage{float}
\usepackage[dvipsnames]{xcolor}
\usepackage[colorlinks=true,linkcolor=black,citecolor=black,urlcolor=black]{hyperref}
\graphicspath{{figures/}}

\newcommand{\avg}[1]{\left\langle #1\right\rangle}
\newcommand{\rhos}{\rho_s}
\newcommand{\kx}{S_x^\star}
\newcommand{\ky}{S_y^\star}
\newcommand{\kd}{S_d^\star}

\newcommand{\Vtwo}{V_2}
\newcommand{\Wsix}{W_6}
\newcommand{\figfallback}[1]{%
  \fbox{%
    \begin{minipage}[c][0.42\linewidth][c]{0.92\linewidth}
      \centering
      \small #1
    \end{minipage}%
  }%
}
\newcommand{\maybefig}[2]{%
  \IfFileExists{#1}{\includegraphics[width=\linewidth]{#1}}{\figfallback{#2}}%
}

\begin{document}

\title{One-sided stripe supersolidity from engineered non-axisymmetric dipolar interactions}

\author{Chao Zhang}
\email{chaozhang@ahnu.edu.cn}
\affiliation{Department of Physics, Anhui Normal University, Wuhu, Anhui 241000, China}

\begin{abstract}
A supersolid combines density order with phase coherence, and doped lattice solids ask whether added defects can become coherent without melting the ordered background. We study a soft-core Bose-Hubbard model with isotropic hopping and an engineered non-axisymmetric dipolar interaction, \(V_{ij}=V_2(x_{ij}^2-y_{ij}^2)/r_{ij}^5+W_6/r_{ij}^6\), where the sign-changing \(d_{x^2-y^2}\) component selects a fixed \((q,0)\) stripe channel and the \(W_6/r^6\) core stabilizes the short-distance attractive branch.
Using sign-problem-free quantum Monte Carlo method with worm algorithm, we find that the half-filled stripe parent responds asymmetrically to doping: the hole side forms locked commensurate stripe solids with vanishing superfluid stiffness, whereas the particle side forms a stripe supersolid with finite compressibility \(\kappa>0\), finite superfluid stiffness \(\rho_s>0\), and enhanced double occupancy \(D\).
Keeping the same off-site kernel while increasing \(U/t\) toward the hard-core limit shows that the particle-side supersolid disappears once doublon-like defects are projected out.
Thus the engineered dipolar kernel selects the fixed \((q,0)\) stripe channel, while onsite softness selects the phase-coherent defect sector.
\end{abstract}

\maketitle

\noindent\textbf{Introduction.--}
Supersolidity requires static density order and phase coherence to survive in the same state~\cite{ProkofevSvistunov2005Supersolid,Boninsegni2012Supersolids}.
In lattice bosons, this question has often been posed after a density wave is already stabilized by extended interactions: do doped vacancies or interstitials delocalize without melting the solid, or do they phase separate~\cite{Goral2002,Sengupta2005,Wessel2005,Heidarian2005,BoninsegniProkofev2005Triangular,Dang2008,Danshita2009,Pollet2010,Capogrosso2010,Trefzger2011}?
Soft-core and nonmonotonic long-range interactions further broaden this setting by allowing cluster, quasisolid, and quasisupersolid order~\cite{Saccani2011SoftCore,Zhang2026Quasisolid}.
Those works established doped-solid supersolidity, but they leave open a sharper design question: can the density-ordering channel and the mobility of doped defects be controlled by different ingredients of the Hamiltonian?

Long-ranged dipolar and extended Bose-Hubbard models already provide a broad route to lattice solids and doped supersolids.
The issue here is more specific: the momentum channel of the density order should be fixed by the off-site interaction geometry, while the mobility of doped defects is controlled separately by onsite softness.
Bare or simply tilted dipoles can be anisotropic and support stripe or supersolid tendencies~\cite{Macia2012,Macia2014,Bombin2017,Zhang2015TiltedDipoles,Zhang2022TiltedDipoles3D}, but changing their orientation generally changes the ordering geometry, defect energetics, and short-distance stability together rather than isolating these roles.
Microwave-dressed polar molecules offer a natural route to this separation because microwave coupling between rotational states, together with dc-field and lattice controls, can engineer effective interactions beyond the bare dipole-dipole form~\cite{Baranov2008,Lahaye2009,Bohn2017,Micheli2006,Buchler2007,Micheli2007,Gorshkov2011,Yan2013,Li2021Tuning,Li2023Tunable}.
Recent shielding, field-linked-resonance, molecular-condensate, and droplet experiments show that stable, coherent, strongly interacting molecular gases with microwave-induced interactions are now accessible~\cite{Matsuda2020,Anderegg2021,Schindewolf2022,Chen2023,Bigagli2023,Bigagli2024,Zhang2026Droplets}.
Accordingly, the key ingredient in the model below is not simply a stronger or longer-ranged repulsion.
The off-site kernel combines a sign-changing \(d_{x^2-y^2}\)-like term, \(V_2(x^2-y^2)/r^5\), with a repulsive \(W_6/r^6\) core: the interaction is repulsive along one lattice axis and attractive along the other, while the core counteracts the attractive axial channel at short distances.
The stripe selection therefore comes from the momentum dependence of the off-site interaction kernel itself, not from anisotropic hopping or onsite softness.

\begin{figure}[t!]
\centering
\IfFileExists{figures/fig1_interaction_kernel.pdf}{\includegraphics[width=0.86\linewidth]{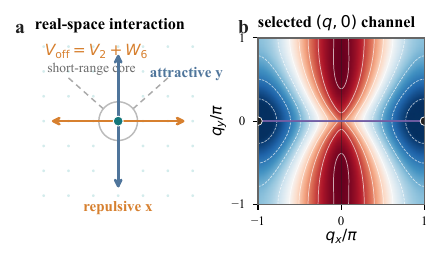}}{\figfallback{Fig. 1: real-space interaction geometry and selected off-site interaction channel.}}
\caption{\textbf{Interaction-selected stripe channel.}
(a) The off-site interaction is repulsive along \(x\), attractive along \(y\), and has a short-range repulsive core.
(b) The momentum-space off-site kernel, \(\widetilde V_{\rm off}(\bm q)\), has a low axial \((q,0)\) channel, while onsite \(U\) does not choose the ordering direction.
In (b), colors show the relative value of \(\widetilde V_{\rm off}(\bm q)\) after subtracting its Brillouin-zone average; the low axial region marks the selected \((q,0)\) channel.
}
\label{fig:model}
\end{figure}

In the hard-core limit this non-axisymmetric kernel selects a fixed \((q,0)\) stripe channel whose leading wave vector changes with filling~\cite{Zhang2026StripeFamily}.
This Letter adds finite onsite repulsion and asks how the selected stripe background responds on its particle and hole sides.
The quantum Monte Carlo results give a strongly asymmetric answer: lowering the chemical potential accesses neighboring commensurate locked stripe solids, while particle doping produces a doped stripe supersolid (DSS), in which stripe order \(\kx\) coexists with compressibility \(\kappa\) and finite superfluid stiffness \(\rho_s\).
The decisive control is that this particle-side DSS disappears when increasing \(U/t\) closes the doublon channel and restores hard-core-like behavior.
Figure~\ref{fig:model} summarizes the interaction-selected part of the mechanism: the dipolar kernel fixes the stripe channel before onsite softness selects which doped defects become phase coherent.

\noindent\textbf{Model and diagnostics.--}
We simulate a soft-core Bose-Hubbard model on an \(L\times L\) square lattice with periodic boundary conditions.
The Hamiltonian is
\begin{equation}
\begin{aligned}
H={}&-t\sum_{\langle i,j\rangle}(b_i^\dagger b_j+\mathrm{H.c.})
+\frac{U}{2}\sum_i n_i(n_i-1)-\mu\sum_i n_i\\
&+\sum_{i<j}\left[
\Vtwo\frac{x_{ij}^2-y_{ij}^2}{r_{ij}^5}
+\frac{\Wsix}{r_{ij}^6}\right]n_i n_j .
\end{aligned}
\label{eq:modelham}
\end{equation}
Here \(n_i=b_i^\dagger b_i\), \(\mu\) is the chemical potential, \(\bm r_{ij}=(x_{ij},y_{ij})\neq(0,0)\) is the lattice displacement used with periodic boundary conditions, \(r_{ij}=|\bm r_{ij}|\), and
\(V_{ij}\equiv \Vtwo(x_{ij}^2-y_{ij}^2)/r_{ij}^5+\Wsix/r_{ij}^6\) denotes the off-site density-density coupling.
We evaluate \(V_{ij}\) using the displacement \(\bm r_{ij}\) defined above, and set \(t=1\) as the energy unit.
For \(\Vtwo>0\), the non-axisymmetric tail is repulsive along the \(x\) axis and attractive along the \(y\) axis, while \(\Wsix>0\) provides a short-range repulsive core.
The kernel is treated as an engineered effective interaction with the symmetry channel motivated by dressed molecules; no hopping anisotropy is imposed.
For the main phase diagram we keep the shape of the interaction fixed, \(U/\Vtwo=30/7\) and \(\Wsix/\Vtwo=3/7\), vary only the overall interaction scale \(\eta=\Vtwo/t\) and the chemical potential, and take the hard-core limit as \(U=\infty\).
The data along \(\eta=3.5\) therefore correspond to \((U,\Vtwo,\Wsix)/t=(15,3.5,1.5)\).
At quadratic order in a weak density modulation, the onsite term shifts all nonzero density-wave momenta equally, whereas the Fourier transform of the off-site interaction, \(\widetilde V_{\rm off}(\bm q)\), supplies the momentum dependence that selects the stripe channel; the derivation is given in the Supplemental Material~\cite{SupplementalMaterial}.

\noindent\textbf{Observables and phase criteria.--}
We use path-integral quantum Monte Carlo with worm algorithm~\cite{Prokofev1998Continuous,Prokofev1998}.
The filling is \(n=\avg{N}/L^2\), where \(N=\sum_i n_i\) is the total particle number, and the doping relative to the half-filled stripe parent is \(\delta n=n-1/2\).
We obtain the compressibility \(\kappa=\beta(\avg{N^2}-\avg{N}^2)/L^2\) and superfluid stiffness from winding-number fluctuations,
\(\rho_{s,\alpha}=\avg{W_\alpha^2}/(\beta t)\), with \(\rho_s=(\rho_{s,x}+\rho_{s,y})/2\)~\cite{Pollock1987}.
Density order is obtained from the equal-time structure factor
\begin{equation}
S(\bm q)=\frac{1}{L^2}\left\langle \left|\sum_j(n_j-n)e^{i\bm q\cdot\bm r_j}\right|^2\right\rangle ,
\end{equation}
where \(\bm q=(2\pi m_x/L,2\pi m_y/L)\) is a nonzero lattice momentum.
We fold momenta into the first quadrant and define \(\kx\) as the maximum nonzero \(S(\bm q)\) in the selected \(x\)-stripe \((q,0)\) channel, where \(q\) denotes the axial component of \(\bm q\).
The transverse \((0,q)\) and diagonal \((q,q)\) channels, denoted \(\ky\) and \(\kd\), are monitored only as checks against competing density-wave orientations.
When these competing channels are shown in a single normalized diagnostic, we use \(S_{\rm comp}^\star\equiv\max(\ky,\kd)\).
The leading folded wave vector \(\bm q_1^{\rm fold}=(q_x^{\rm fold},q_y^{\rm fold})\) is the location of the largest nonzero structure-factor peak.

Onsite softness is measured by
\(D=L^{-2}\sum_i\avg{n_i(n_i-1)/2}\), where \(n_i=b_i^\dagger b_i\), and by occupation probabilities
\(P_{\ge m}=L^{-2}\sum_i\sum_{\ell\ge m}P_i(\ell)\), with \(P_i(\ell)\) the probability that site \(i\) has occupation \(\ell\).
These local observables identify the soft-core defect channel and check for high-occupancy accumulation.
Different phases are determined from \(n(\mu)\), compressibility \(\kappa\), superfluid stiffness \(\rho_s\), and selected-stripe peak \(\kx\).
Stripe-ordered plateaus with finite \(\kx\), suppressed \(\kappa\), and vanishing \(\rho_s\) are locked stripe solids.
The half-filled locked parent is denoted HSS, while LSS refers to neighboring non-half-filled locked stripe solids.
A doped stripe supersolid (DSS) is non-plateau and has finite \(\kx\), \(\kappa\), and \(\rho_s\); the directional stiffness components are anisotropic, \(\rho_{s,x}>\rho_{s,y}>0\).
Uniform superfluids have finite \(\rho_s\) and \(\kappa\) but no dominant selected-stripe peak \(\kx\).
Further convergence checks and auxiliary diagnostics are given in the Supplemental Material~\cite{SupplementalMaterial}.

\begin{figure}[t!]
\centering
\includegraphics[width=\linewidth]{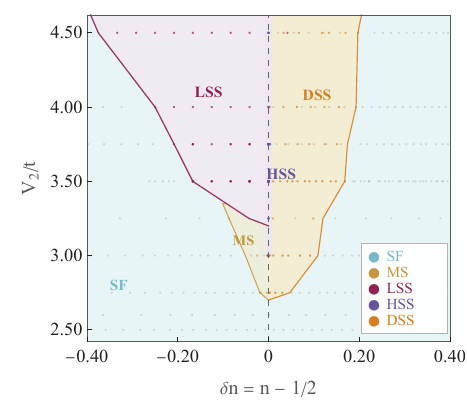}
\caption{\textbf{Fixed-shape phase diagram.}
Phase assignments in the \((\delta n,\eta)\) plane, with \(\delta n=n-1/2\), \(\eta=\Vtwo/t\), and fixed ratios \(U/\Vtwo=30/7\), \(\Wsix/\Vtwo=3/7\).
Small gray dots denote simulated points; colored points and domains summarize the identified phases: superfluid (SF), mobile stripe (MS), locked stripe solid (LSS), half-filled stripe solid (HSS), and doped stripe supersolid (DSS).
The HSS forms the central parent at \(\delta n=0\).
}
\label{fig:phase}
\end{figure}

\begin{figure}[t!]
\centering
\makebox[\linewidth][c]{\includegraphics[width=1.06\linewidth]{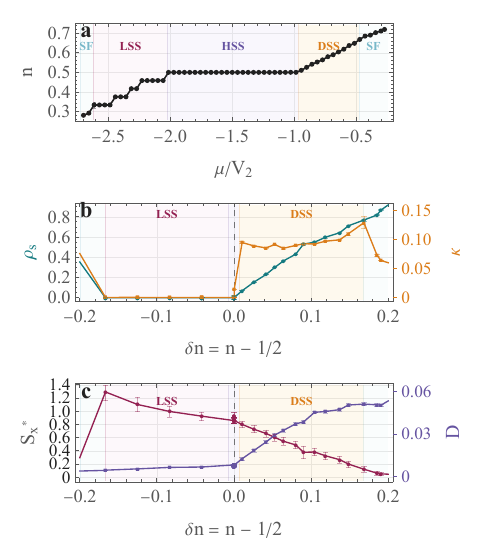}}
\caption{\textbf{Particle-hole-asymmetric stripe response along \(\eta=\Vtwo/t=3.5\).}
All panels show \(L=24\) results at \((U,\Vtwo,\Wsix)/t=(15,3.5,1.5)\), i.e., the data set along the \(\eta=3.5\) line of the fixed-shape parameter trajectory in Fig.~\ref{fig:phase}.
(a) The \(n(\mu)\) curve shows a broad half-filled stripe parent and asymmetric behavior on the two sides.
(b) Relative to half filling, the left and right axes show \(\rho_s\) and \(\kappa\), respectively: the hole side remains locked, whereas particle doping produces a compressible state with \(\kappa>0\) and finite superfluid stiffness \(\rho_s>0\).
(c) The left axis shows that the selected-stripe structure-factor weight \(\kx\) remains dominant across the same interval, while the right axis shows the onsite pair density \(D\).
Error bars denote statistical uncertainties.
}
\label{fig:cut}
\end{figure}

\noindent\textbf{Phase diagram.--}
Figure~\ref{fig:phase} first shows the fixed-shape phase diagram in the \((\delta n,\eta)\) plane, obtained at \(U/\Vtwo=30/7\) and \(\Wsix/\Vtwo=3/7\) while varying \(\mu\).
We call the half-filled stripe solid at \(\delta n=0\) the parent state; its particle and hole sides are strongly asymmetric.
For \(\eta\lesssim2.55\) the system is SF. On the hole side, for \(2.6\lesssim\eta\lesssim3.1\) stripe order \(\kx\) remains mobile with finite \(\rho_s\); for \(3.1\lesssim\eta\lesssim3.3\), decreasing \(\delta n\) gives locked stripes with suppressed \(\kappa,\rho_s\), mobile stripes with finite \(\rho_s\), then SF; and for \(\eta\gtrsim3.3\), the hole-side stripe region identified in the scan is locked until it reaches SF.
The particle side instead retains \(\kx\), \(\kappa\), and \(\rho_s\) over a finite window of $\delta n$.
The use of \(\delta n\), rather than \(n\) alone, emphasizes that the comparison is made around the same half-filled stripe parent.
The key information in Fig.~\ref{fig:phase} is that the same interaction-selected \((q,0)\) channel responds asymmetrically on the two sides when $\eta \gtrsim3.3$.
On the hole side, commensurability locks stripe solids with finite \(\kx\) but suppressed \(\kappa,\rho_s\); on the particle side, the stripe state can keep \(\kappa>0\) and \(\rho_s>0\).

\noindent\textbf{One-sided supersolidity.--}
Figure~\ref{fig:cut} gives the most direct view of this asymmetry along \(\eta=3.5\).
Below the half-filled parent, the density locks into neighboring commensurate stripe plateaus rather than evolving continuously; \(\kx\) remains large, while \(\kappa\) and \(\rho_s\) vanish.
For such plateau states, finite-size checks are most naturally made at a chemical potential inside the plateau, because different finite periodic lattices can accommodate slightly different commensurate stripe fillings.
Particle doping instead gives continuous density evolution, finite \(\kappa\), finite \(\rho_s\), and persistent dominant \(\kx\).
The onsite pair weight \(D\) distinguishes the two sides of the parent.
In the half-filled parent and in the hole-side LSS, where \(\rho_s\) is negligible, \(D\) stays close to the background value \(D\simeq7\times10^{-3}\).
On the particle side, where the DSS appears, \(D\) rises to about \(4\times10^{-2}\) near \(\delta n\simeq0.080\) and continues to increase with particle doping, showing that the coherent response is accompanied by the opening of the soft-core doublon channel.
This is the one-sided DSS: \(\kx\), \(\kappa\), and \(\rho_s\) coexist only for \(\delta n>0\).
The onsite pair density \(D\) is not an order parameter for the DSS; it is a doublon-sensitive diagnostic that tracks the soft-core channel accompanying \(\rho_s\).
The key point is that \(\kx\) remains large on both sides of the parent.
Thus the asymmetry is not a switch of density-wave orientation: the hole side forms locked \((q,0)\) stripe plateaus, whereas particle doping makes the same stripe channel phase coherent.

\noindent\textbf{Soft-core control.--}
For finite-\(U\) Bose-Hubbard models, particle and hole dopings are not related by an exact microscopic symmetry in general; their excitation gaps and dispersions are inequivalent away from special particle-hole-symmetric limits~\cite{Fisher1989,FreericksMonien1996}.
Thus an asymmetry between \(\delta n>0\) and \(\delta n<0\) is not, by itself, the central result.
The decisive question is whether the particle-side DSS, with \(\kx\), \(\kappa\), and \(\rho_s\), survives when the soft-core doublon channel is closed.
In the hard-core limit, \(n_i=1-n_i^h\) maps the density interaction back to the same \(V_{ij}n_i^h n_j^h\) form, up to a shifted chemical potential and constants.
This transformation only relabels particles as holes on the same square lattice; it does not include an additional \(x\leftrightarrow y\) rotation.
Therefore the off-site kernel \(V_{ij}\), including its \(d_{x^2-y^2}\) sign pattern, is unchanged in the hard-core hole Hamiltonian.
A separate \(x\leftrightarrow y\) operation would instead describe a different, rotated interaction profile.
Thus the hard-core baseline gives the proper reference for whether a one-sided DSS is genuinely caused by softness.

\begin{figure}[t!]
\centering
\includegraphics[width=0.98\columnwidth]{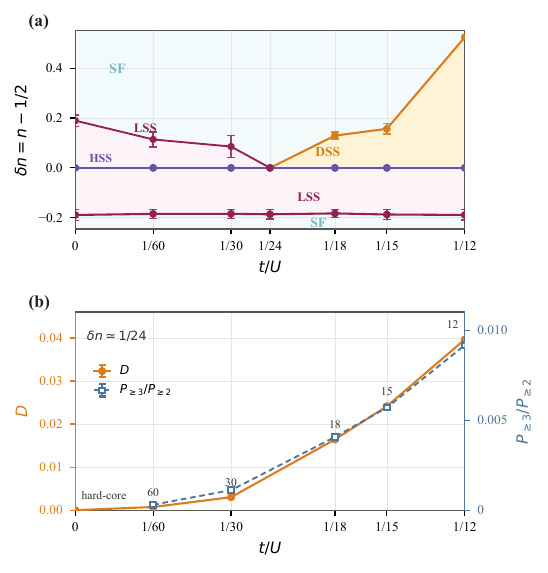}
\caption{\textbf{Soft-core-to-hard-core control.}
(a) At fixed \(\Vtwo/t=3.5\) and \(\Wsix/t=1.5\), the half-filled HSS parent remains at \(\delta n=0\), while the particle-side DSS window appears for \(U/t=12,15,18\) but is absent for \(U/t=24,30,60\) and hard-core.
Error bars in (a), shown only on the plotted phase-window boundaries, denote the boundary-location uncertainty set by adjacent sampled filling sectors. 
(b) At sampled particle-side sectors near \(\delta n\simeq1/24\), the left axis shows the onsite pair weight \(D=L^{-2}\sum_i\avg{n_i(n_i-1)/2}\), while the right axis shows the higher-occupancy ratio \(P_{\ge3}/P_{\ge2}\).
Error bars in (b) denote statistical uncertainties.
}
\label{fig:softnessControl}
\end{figure}

Finite \(U\) changes the low-energy Hilbert space.
In a strong-coupling stripe background \(\{n_i^0\}\), the local costs of adding and removing one boson at \(t=0\) are
\begin{equation}
\epsilon_i^+=Un_i^0-\mu+\Phi_i,\qquad
\epsilon_i^-=\mu-U(n_i^0-1)-\Phi_i ,
\end{equation}
with \(\Phi_i=\sum_{j\ne i}V_{ij}n_j^0\).
On an occupied stripe site, the particle defect is doublon-like and costs \(U-\mu+\Phi_i\); the hole defect instead removes density from the stripe background and has no negative-occupation counterpart.
Equivalently, the dilute one-defect Hamiltonians have the schematic form
\begin{equation}
\begin{aligned}
H_{\rm def}^+&=\sum_i\epsilon_i^+p_i^\dagger p_i
-\sum_{\langle i,j\rangle}t_{ij}^+(p_i^\dagger p_j+{\rm H.c.})+\cdots,\\
H_{\rm def}^-&=\sum_i\epsilon_i^-h_i^\dagger h_i
-\sum_{\langle i,j\rangle}t_{ij}^-(h_i^\dagger h_j+{\rm H.c.})+\cdots,
\end{aligned}
\end{equation}
where \(p_i^\dagger\) and \(h_i^\dagger\) create particle and hole defects on the stripe background, and the ellipses denote defect-defect interactions and higher-order processes.
The two Hamiltonians are not particle-hole partners at finite \(U\): their onsite landscapes \(\epsilon_i^\pm\) differ, and their hopping matrix elements scale with different bosonic factors, \(t_{ij}^+\sim t\sqrt{(n_i^0+1)(n_j^0+1)}\) and \(t_{ij}^-\sim t\sqrt{n_i^0n_j^0}\).
The relevant hard-core crossover scale is therefore the effective doublon gap
\(\Delta_2=\min_i(U-\mu+\Phi_i)\).
Hard-core-like behavior is recovered only when \(\Delta_2\) is large compared with the low-energy scales of the stripe state, such as hopping and temperature, so that the \(n_i\ge2\) sector is projected out.
Figure~\ref{fig:softnessControl}(a) supplies this control directly: at fixed \(\Vtwo/t=3.5\) and \(\Wsix/t=1.5\), the particle-side DSS window is present for \(U/t=12,15,18\), but is absent for \(U/t\ge24\) and in the hard-core limit.
In the displayed comparison shown in Fig.~\ref{fig:softnessControl}(b), the onsite pair weight \(D\) grows as \(t/U\) is increased and falls toward its hard-core value as \(t/U\to0\), while the high-occupancy ratio \(P_{\ge3}/P_{\ge2}\) remains below \(10^{-2}\).
The finite-\(U\) channel that tracks the DSS is therefore dominated by doublon-like fluctuations, not by higher onsite accumulation.
The one-sided DSS, defined by coexisting \(\kx\), \(\kappa\), and \(\rho_s\), is therefore not just a generic finite-\(U\) asymmetry; it disappears when the doublon sector is projected out.

\begin{figure}[t!]
\centering
\maybefig{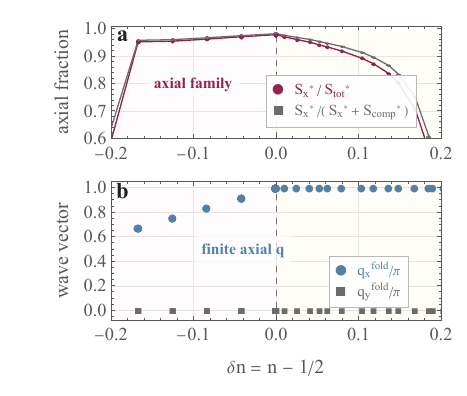}{Fig. 5: selected-stripe order and leading folded wave-vector diagnostics.}
\caption{\textbf{Selected-stripe and leading-wave-vector diagnostics.}
These results use the \(L=24\), \(\eta=3.5\) fixed-shape cut, \((U,\Vtwo,\Wsix)/t=(15,3.5,1.5)\), also shown in Fig.~\ref{fig:cut}.
Normalized channel weights and the leading folded wave vector show that the ordered states are carried by the same interaction-selected \((q,0)\) stripe channel.
Here \(S_{\rm comp}^\star=\max(\ky,\kd)\) denotes the larger of the transverse and diagonal competing-channel peaks.
The plotted \(q_x^{\rm fold}\) and \(q_y^{\rm fold}\) are the two components of \(\bm q_1^{\rm fold}\).
The dominant structure-factor weight remains in the \(x\)-stripe channel, while \(q_x^{\rm fold}\) drifts with filling and \(q_y^{\rm fold}\) remains near zero.
}
\label{fig:qfamily}
\end{figure}

\noindent\textbf{Selected stripe channel and checks.--}
The relevant check is whether the finite-stiffness particle-side state retains dominant interaction-selected stripe order.
Figure~\ref{fig:qfamily} shows this directly: the dominant density peak \(\kx\) stays in the selected \(x\)-stripe \((q,0)\) channel, \(q_x^{\rm fold}\) changes smoothly with filling, and \(q_y^{\rm fold}\) remains near zero.
Long-wavelength density and local-occupation diagnostics give additional support for this assignment.
The smallest-momentum density weight remains orders of magnitude below \(\kx\) across the DSS window, consistent with finite-\(q\) stripe order without a visible macroscopic density-segregation signal.
High-occupation events remain rare, so the increase of \(D\) is mainly doublon-like soft-core weight rather than a buildup of multiply occupied sites.
Finally, the particle-side DSS has anisotropic but two-dimensional phase coherence, with \(\rho_{s,x}>\rho_{s,y}>0\), as expected in a stripe-ordered state.
Finite-size checks, a representative \(\beta\)-convergence check, real-space diagnostics, and the numerical bounds for these stability checks are in the Supplemental Material~\cite{SupplementalMaterial}.

\noindent\textbf{Experimental connection.--}
In a microwave-dressed polar-molecule implementation, a two-dimensional optical lattice sets \(t\), while rotational-state dressing with microwave and dc fields engineers off-site couplings beyond the statically polarized \(1/r^3\) dipole-dipole interaction~\cite{Micheli2006,Buchler2007,Micheli2007,Gorshkov2011}.
The onsite \(U\) is the energy cost of double occupation in one Wannier orbital and is controlled by confinement and local molecular interactions. Short-range collisional stability can be enhanced by electric or microwave shielding~\cite{Matsuda2020,Anderegg2021,Schindewolf2022,Bigagli2023}, while field-linked resonances provide a separate route for tuning molecular interactions~\cite{Chen2023}.
The experimental target is the phase diagram itself: \(\eta=\Vtwo/t\) can be changed by tuning the interaction-to-hopping scale, while \(\delta n\) can be tuned by changing the total molecule number or read out locally in a smooth trap.
The required observables are standard cold-gas diagnostics~\cite{Bloch2008,Gross2021QuantumMicroscopy}: density profiles and fluctuations give \(n(\mu)\) and \(\kappa\), density correlations, coherent light scattering, or noise correlations give \(S(\bm q)\), and phase-coherence probes access the superfluid response. Number-resolved doublon imaging probes \(D\)~\cite{Hartke2020DoublonHole}.
In this protocol, HSS/LSS are identified by stripe order with density locking and suppressed \(\kappa,\rho_s\), whereas DSS has the same selected-stripe order together with continuous \(n\), finite \(\kappa,\rho_s\), and enhanced \(D\).
In a smooth trap, the local density varies across the cloud. The high-density side of the half-filled stripe region can support a DSS, whereas the low-density side passes through commensurate LSS plateaus, producing an asymmetric spatial phase sequence.

\noindent\textbf{Conclusion.--}
We have shown that a non-axisymmetric dipolar kernel and finite onsite softness play separate, experimentally distinguishable roles in doped stripe order.
The engineered off-site kernel selects the axial \((q,0)\) stripe channel, while finite \(U\) controls which defects become mobile within that channel.
This produces a one-sided DSS: particle doping opens a doublon-assisted coherent stripe state, whereas the hole side consists of commensurate locked stripe plateaus.
Because the DSS disappears as \(U/t\) is increased toward the hard-core limit, the central signature is the tunable loss of the particle-side supersolid when the doublon channel is closed.
Engineered dipolar molecules therefore offer a route to use onsite Hilbert-space softness as a control knob for defect-mediated supersolidity inside a fixed interaction-selected stripe channel.

\noindent\textbf{Acknowledgments.--}
C.Z. thanks Nikolay Prokof'ev for helpful discussions.
C.Z. acknowledges support from the National Natural Science Foundation of China under Grant No.~12204173.

\bibliography{references}

\clearpage
\onecolumngrid
\pagestyle{plain}
\begingroup
\setcounter{section}{0}
\setcounter{equation}{0}
\setcounter{figure}{0}
\setcounter{table}{0}
\renewcommand{\thefigure}{S\arabic{figure}}
\renewcommand{\thetable}{S\arabic{table}}
\renewcommand{\theHsection}{supp.\arabic{section}}
\renewcommand{\theHequation}{supp.\arabic{equation}}
\renewcommand{\theHfigure}{supp.\arabic{figure}}
\renewcommand{\theHtable}{supp.\arabic{table}}
\def\ARXIVCOMBINED{1}
\vspace*{-2.6em}
\begin{center}
{\large\bf Supplemental Material for ``One-sided stripe supersolidity from engineered non-axisymmetric dipolar interactions''\par}
\vspace{0.15em}
{\normalsize Chao Zhang\par}
\vspace{0.15em}
{\it Department of Physics, Anhui Normal University, Wuhu, Anhui 241000, China\par}
\end{center}
\vspace{0.35em}

\section{Real-space diagnostics}

Density order is measured by the equal-time structure factor
\begin{equation}
S(\bm q)=\frac{1}{L^2}\left\langle\left|\sum_j(n_j-n)e^{i\bm q\cdot\bm r_j}\right|^2\right\rangle,
\qquad
\bm q=\left(\frac{2\pi m_x}{L},\frac{2\pi m_y}{L}\right),
\end{equation}
where \(\bm q\) is a nonzero lattice momentum.
We fold momenta into the first quadrant and define \(\kx\) as the maximum nonzero \(S(\bm q)\) in the selected \(x\)-stripe \((q,0)\) channel, where \(q\) denotes the axial component of \(\bm q\).
Phase coherence is measured from the superfluid stiffness computed from winding-number fluctuations,
\begin{equation}
\rho_{s,\alpha}=\frac{\langle W_\alpha^2\rangle}{\beta t},\qquad
\rho_s=\frac{\rho_{s,x}+\rho_{s,y}}{2},
\end{equation}
and compressibility from number fluctuations,
\begin{equation}
\kappa=\frac{\beta}{L^2}\left(\langle N^2\rangle-\langle N\rangle^2\right).
\end{equation}
Local occupation statistics are used to identify the soft-core channel and to distinguish controlled doublon weight from local high-occupation accumulation:
\begin{equation}
D=\frac{1}{L^2}\sum_i\left\langle\frac{n_i(n_i-1)}{2}\right\rangle,\qquad
P_{\ge m}=\frac{1}{L^2}\sum_i\sum_{\ell\ge m}P_i(\ell),
\end{equation}
where \(P_i(\ell)\) is the probability of finding \(\ell\) bosons on site \(i\).
To check that the enhanced onsite pair weight \(D\) reflects a controlled doublon-like soft-core channel, rather than local high-occupation accumulation, we monitor \(P_{\ge3}/P_{\ge2}\) and \(n_{\rm loc}^{\rm max}=\max_i\langle n_i\rangle\).
At a particle-side DSS point, \(\mu/t=-2.6\), we find \(P_{\ge3}/P_{\ge2}\simeq0.007\) and \(n_{\rm loc}^{\rm max}\simeq1.08\).
Thus the increase of \(D\) is dominated by \(n_i=2\) occupation, not by a buildup of \(n_i\ge3\) states.
This check is distinct from the superfluid-stiffness anisotropy: in a stripe-ordered phase \(\rho_{s,x}\) and \(\rho_{s,y}\) need not be equal, but both being finite indicates phase coherence in both spatial directions.
Real-space maps are therefore used only to illustrate the stripe translation pattern retained during Monte Carlo sampling; different phases are determined from \(S(\bm q)\), \(\rho_s\), \(\kappa\), and the local occupation statistics above.
\pagebreak[3]
Figure~\ref{figS:maps} shows representative density and onsite pair-density maps on the \(\eta=\Vtwo/t=3.5\) cut.
The particle-doped state retains the axial stripe pattern while developing much stronger onsite pair weight, consistent with mobile soft-core defects on the stripe background.

\begin{figure}[H]
\centering
\includegraphics[width=0.88\linewidth,height=0.56\textheight,keepaspectratio]{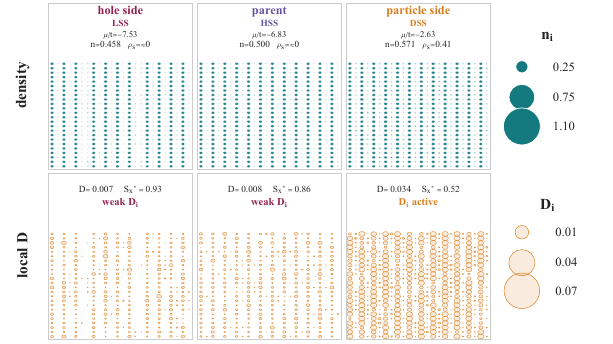}
\caption{
\textbf{Real-space density maps and soft-core defect maps.}
All panels show \(L=24\) results at \((U,\Vtwo,\Wsix)/t=(15,3.5,1.5)\).
Time-averaged density maps show the local density \(n_i\) and onsite pair density \(D_i=\avg{n_i(n_i-1)/2}\) for a hole-side LSS, the half-filled stripe parent, and a particle-doped stripe state.
}
\label{figS:maps}
\end{figure}

\section{Momentum-space selection of the stripe channel}

This section explains the organizing statement used in the Letter: the off-site interaction selects the stripe channel, whereas the onsite softness controls which doped defects can move in that channel.
The off-site part of the Hamiltonian can be written as
\begin{equation}
H_{\rm off}=\frac{1}{2}\sum_{i\neq j}V(\bm r_i-\bm r_j)n_i n_j,
\qquad
V(\bm r)=\Vtwo\frac{x^2-y^2}{r^5}+\frac{\Wsix}{r^6},
\end{equation}
which is equivalent to the \(i<j\) form in the main text.
Consider a weak density modulation around a uniform background,
\begin{equation}
n_i=n+\delta n_i,\qquad
\delta n_i=\sum_{\bm q}\delta n_{\bm q}e^{i\bm q\cdot\bm r_i},
\qquad
\delta n_{\bm q}=\frac{1}{L^2}\sum_i\delta n_i e^{-i\bm q\cdot\bm r_i}.
\end{equation}
The \(\bm q=0\) component changes only the average density and is fixed once the filling is fixed.
For nonzero \(\bm q\), the quadratic change of the off-site interaction energy is
\begin{equation}
\frac{\delta E_{\rm off}}{L^2}
=
\frac{1}{2}\sum_{\bm q\neq0}
\widetilde V_{\rm off}(\bm q)|\delta n_{\bm q}|^2,
\qquad
\widetilde V_{\rm off}(\bm q)=\sum_{\bm r\neq0}V(\bm r)e^{i\bm q\cdot\bm r}.
\end{equation}
Thus \(\widetilde V_{\rm off}(\bm q)\) is the momentum-space quadratic coefficient for density waves generated by the off-site interaction alone.
Equivalently, among weak density modulations with the same amplitude, the off-site interaction favors wave vectors where \(\widetilde V_{\rm off}(\bm q)\) is smallest.

The onsite term is different because it is local:
\begin{equation}
H_U=\frac{U}{2}\sum_i n_i(n_i-1).
\end{equation}
Expanding the same weak modulation gives a \(q\)-independent quadratic contribution,
\begin{equation}
\frac{\delta E_U}{L^2}
=\frac{U}{2}\sum_{\bm q\neq0}|\delta n_{\bm q}|^2 .
\end{equation}
Therefore the interaction contribution to the quadratic density kernel is
\begin{equation}
K_{\rm int}(\bm q)=U+\widetilde V_{\rm off}(\bm q).
\end{equation}
This notation emphasizes that the expression is the interaction part of the density-mode cost; hopping and quantum fluctuations enter the full many-body response separately.
The important point is that \(U\) shifts all density-wave momenta by the same amount.
It can make density modulation more or less costly overall, but it does not choose between \((q,0)\), \((0,q)\), and \((q,q)\) density waves.
That choice comes from the momentum dependence of \(\widetilde V_{\rm off}(\bm q)\).

For the mixed \(d_{x^2-y^2}+1/r^6\) kernel used here, the smallest values of \(\widetilde V_{\rm off}\) occur in the axial \((q,0)\) channel.
This is why the ordered states are analyzed by the leading peak in the selected \((q,0)\) sector, while the transverse \((0,q)\) and diagonal \((q,q)\) sectors are monitored as competing channels.
The value of the leading \(q\) can change with filling because different commensurate stripe spacings fit different densities, but the orientation channel remains the same.

Finite onsite softness enters at the next stage, after the off-site geometry has selected the stripe channel.
Changing the particle number probes inequivalent particle- and hole-like excitations. Added particles can enter the local \(n\ge2\) sector and become mobile soft-core defects. Removing a particle instead creates a vacancy with no analogous \(n\ge2\) channel; in the present phase diagram, lowering \(\mu\) locks the density into neighboring commensurate stripe plateaus rather than producing a continuously hole-doped phase.
This is the sense in which the interaction geometry selects the \((q,0)\) stripe channel, while onsite softness selects the mobile defect sector.

\section{Particle-hole transformation and hard-core baseline}

A finite-\(U\) Bose-Hubbard model is not generically particle-hole symmetric: particle and hole excitations have different energetic and kinetic structure except in special symmetric limits or at emergent low-energy particle-hole-symmetric critical points~\cite{Fisher1989,FreericksMonien1996}.
Therefore the central claim of the Letter is not the existence of particle-hole asymmetry alone.
The useful baseline is the hard-core limit.

This limit removes the \(n\ge2\) onsite sector while keeping the same off-site interaction geometry, and therefore asks a sharper control question: if the stripe-selecting kernel is unchanged but doublon-like defects are forbidden, does the particle-side DSS survive?
Because the hard-core particle-hole transformation maps the model back to the same density-interaction form, up to a chemical-potential shift, the hard-core comparison isolates onsite softness rather than changing the selected stripe channel.

For hard-core bosons, write \(n_i=1-n_i^h\) and \(b_i^\dagger\to h_i\), where \(h_i^\dagger\) creates a hole.
The hopping term keeps the same form, while the density interaction transforms as
\begin{equation}
V_{ij}n_i n_j
=V_{ij}(1-n_i^h)(1-n_j^h)
=V_{ij}-V_{ij}(n_i^h+n_j^h)+V_{ij}n_i^h n_j^h .
\end{equation}
With periodic boundary conditions and no trap or disorder, every site has the same set of relative displacement vectors.
Hence \(V_0=\sum_{j\neq i}V_{ij}\) is site independent, and the linear terms generated by the interaction combine into a uniform hole chemical-potential shift.
The hard-core Hamiltonian becomes
\begin{equation}
H_{\rm hc}
\to
E_0
-t\sum_{\langle i,j\rangle}(h_i^\dagger h_j+{\rm H.c.})
-(V_0-\mu)\sum_i n_i^h
+\sum_{i<j}V_{ij}n_i^h n_j^h .
\end{equation}
Thus particle-hole conjugation leaves the off-site kernel \(V_{ij}\) unchanged and only shifts the chemical potential to \(\mu_h=V_0-\mu\).
In particular, the \(d_{x^2-y^2}\)-like component \(V_2(x^2-y^2)/r^5\) does not change sign under particle-hole conjugation.
An additional spatial interchange \(x\leftrightarrow y\) would change \(x^2-y^2\to-(x^2-y^2)\), but that is a separate rotation of the interaction kernel, not the particle-hole map.

At finite \(U\), the exact mapping fails because particle defects can access the \(n=2\) onsite sector while hole defects cannot access negative occupation.
In a stripe background \(\{n_i^0\}\), the single-defect energy costs, for sites where the corresponding operation is allowed, are
\begin{equation}
\epsilon_i^+=Un_i^0-\mu+\sum_{j\neq i}V_{ij}n_j^0,\qquad
\epsilon_i^-=\mu-U(n_i^0-1)-\sum_{j\neq i}V_{ij}n_j^0 .
\end{equation}
The expression for \(\epsilon_i^-\) applies only to sites with \(n_i^0>0\), since an empty site cannot host an additional hole defect.
The doublon-like soft-core channel is controlled by
\begin{equation}
\Delta_2=\min_i\left(U-\mu+\Phi_i\right),\qquad
\Phi_i=\sum_{j\neq i}V_{ij}n_j^0 ,
\end{equation}
for sites with \(n_i^0=1\).
Strictly, the hard-core model is recovered only at \(U=\infty\).
At finite \(U\), hard-core-like behavior is a crossover expected when \(\Delta_2\) is large compared with the low-energy scales of the stripe state, such as hopping and temperature.
Consistent with this crossover picture, the main-text soft-core control shows a finite particle-side DSS for \(U/t=12,15,18\), but no DSS for \(U/t=24,30,60\) or in the hard-core limit at fixed \(V_2/t=3.5\) and \(W_6/t=1.5\).

\section{Finite-size and beta-convergence checks}

Figure~\ref{figS:finiteL} shows representative finite-size checks for \((U,\Vtwo,\Wsix)/t=(15,3.5,1.5)\).
The HSS and DSS curves use matched fixed filling factors.
For the LSS check, we instead hold \(\mu/t=-7.70\) inside the locked plateau.
This is the natural finite-size protocol for an incompressible plateau: within the plateau, changing \(\mu\) does not continuously change \(n\), and the phase is identified by density locking together with stripe order and vanishing superfluid stiffness.
On finite periodic lattices, however, only certain stripe periods and commensurate fillings fit each system size.
As \(L\) changes, the same plateau can therefore be represented by slightly different allowed fillings.

\begin{figure}[tbp]
\centering
\includegraphics[width=0.72\linewidth,height=0.50\textheight,keepaspectratio]{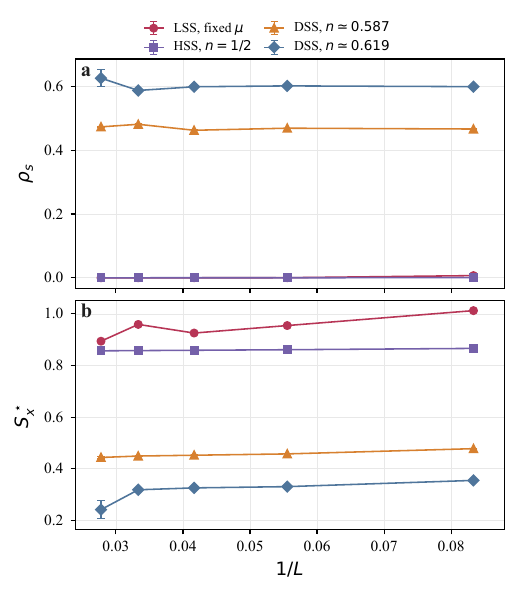}
\caption{
\textbf{Finite-size check at \(\eta=\Vtwo/t=3.5\).}
The superfluid stiffness and selected-stripe structure factor are shown versus \(1/L\) for representative locked stripe solid and particle-doped DSS sectors.
The particle-doped DSS sectors retain finite \(\rhos\) and \(\kx\), while the locked sectors remain non-superfluid within the same analysis protocol.
}
\label{figS:finiteL}
\end{figure}

As a finite-temperature check, we examined a representative particle-doped DSS point, \(L=24\), \((U,\Vtwo,\Wsix)/t=(15,3.5,1.5)\), and \(\mu/t=-2.25\), for several inverse temperatures.
Table~\ref{tabS:beta} compares the matched-filling data sets.
Across this range, the superfluid stiffness remains finite, the selected-stripe peak remains dominant, and the onsite pair weight remains enhanced.
The phase diagnostics are stable over this \(\beta\) range; in practice, \(\beta=24\) is sufficient for the observables reported in the Letter.

\begin{table}[!htbp]
\caption{
\textbf{\(\beta\)-convergence check.}
The representative particle-side DSS point is \(L=24\), \((U,\Vtwo,\Wsix)/t=(15,3.5,1.5)\), and \(\mu/t=-2.25\); numbers in parentheses denote standard errors.}
\label{tabS:beta}
\centering
\scriptsize
\setlength{\tabcolsep}{4pt}
\begin{tabular}{ccccc}
\hline
\(\beta t\) & \(n\) & \(\rhos\) & \(\kx\) & \(D\)\\
\hline
12 & 0.60485 & 0.54215(2) & 0.38078(1) & 0.04361(9)\\
16 & 0.60506 & 0.54211(4) & 0.38037(1) & 0.04359(7)\\
20 & 0.60549 & 0.55928(4) & 0.37746(3) & 0.04342(7)\\
24 & 0.60492 & 0.54609(3) & 0.38291(3) & 0.04336(6)\\
\hline
\end{tabular}
\end{table}

\section{Soft-core-to-hard-core control details}

Figure~4(a) of the main text shows the corresponding phase-diagram evolution with \(t/U\).
At fixed \(\Vtwo/t=3.5\) and \(\Wsix/t=1.5\), the DSS window is finite for \(U/t=12,15,18\), but is absent for \(U/t\ge24\) and in the hard-core limit.
The \(U/t\geq24\) and hard-core comparisons show no coexistence between stripe order and superfluid stiffness on the particle side: the stripe branch either remains locked or gives way directly to a uniform superfluid.
\begin{table}[!h]
\caption{
\textbf{Soft-core-to-hard-core trend at fixed \(\Vtwo/t=3.5\), \(\Wsix/t=1.5\).}
The table summarizes how the particle-side stripe branch changes as the onsite soft-core sector is suppressed.
}
\label{tabS:softness}
\centering
{\renewcommand{\arraystretch}{1.18}
\setlength{\tabcolsep}{0pt}
\small
\begin{tabular}{@{}c@{\hspace{1.4em}}c@{\hspace{1.6em}}l@{\hspace{1.6em}}l@{}}
\hline\hline
\textbf{\(U/t\)} &
\textbf{DSS} &
\textbf{Stripe response} &
\textbf{Soft-core role}\\
\hline
12, 15, 18 &
present &
selected stripe \(+\) finite \(\rho_s\) &
mobile defects\\
\hline
24 &
absent &
small-\(\delta n\) stripe locked &
too weak for DSS\\
\hline
30, 60 &
absent &
no stripe-\(\rho_s\) coexistence &
hard-core-like\\
\hline
\(\infty\) &
absent &
same off-site stripe kernel &
\(n_i\ge2\) states forbidden\\
\hline\hline
\end{tabular}}
\end{table}

Figure~\ref{figS:doublonCurves} gives the corresponding microscopic diagnostic across the particle-doped side.
The onsite pair weight \(D\) quantifies the admixture of the local \(n\ge2\) sector.
The curves show the same hierarchy as the main-text near-\(\delta n=1/24\) check: \(D\) is strictly zero in the hard-core limit, remains small for \(U/t=60,30\), is also small along the \(U/t=24\) curve, and grows rapidly for \(U/t=18,15,12\).
Thus the disappearance of the particle-side DSS window tracks the closure of the finite-\(U\) doublon-like channel over the whole particle-side range, not only at one selected filling.

\begin{figure}[tbp]
\centering
\includegraphics[width=0.78\linewidth]{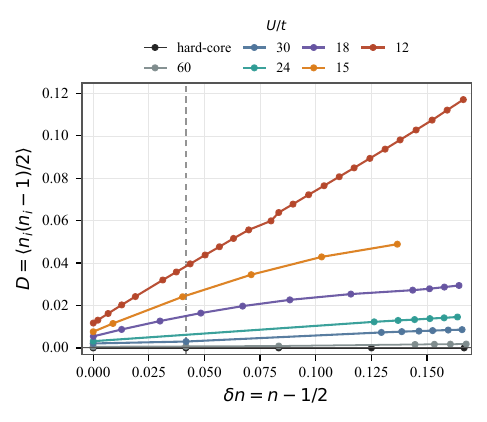}
\caption{
\textbf{Particle-side doublon-channel curves.}
The onsite pair weight \(D=L^{-2}\sum_i\langle n_i(n_i-1)/2\rangle\) is shown versus particle-side doping \(\delta n=n-1/2\) for the same \(U/t\) data sets used in the main-text softness control.
The vertical dashed line marks \(\delta n=1/24\), the particle-side doping used for the comparison in Fig.~4(b) of the main text.
The hard-core curve is identically zero by construction, while finite-\(U\) curves quantify how strongly the local \(n\ge2\) channel is admixed.
}
\label{figS:doublonCurves}
\end{figure}

\section{Experimental basis, implementation, and detection}

The Hamiltonian in the Letter should be understood as an effective lowest-band lattice model: the optical lattice and microwave dressing renormalize the microscopic molecular interaction into the parameters \(t\), \(U\), and \(V_{ij}\).
These parameters describe the dressed and Wannier-projected lattice model.
Dipolar many-body Hamiltonians have already been implemented in several complementary ways.
Magnetic atoms such as Cr, Dy, and Er realize direct magnetic dipole-dipole interactions whose anisotropy is controlled by the quantization axis~\cite{Griesmaier2005,Lahaye2009,Lu2011,Aikawa2012}; optical-lattice experiments with Er further demonstrated an extended Bose-Hubbard model with measurable off-site dipolar couplings~\cite{Baier2016}.
Polar molecules provide a larger electric dipole moment and internal-state control.
Lattice-confined KRb experiments demonstrated dipolar spin exchange~\cite{Yan2013}, while later molecular work showed tunable dipolar interactions and itinerant spin dynamics~\cite{Li2021Tuning,Li2023Tunable}.
The effective interaction studied here is motivated by rotational-state dressing of polar molecules: static electric and microwave fields can tune the strength, sign, and angular dependence of long-range molecular interactions~\cite{Micheli2006,Buchler2007,Micheli2007,Gorshkov2011}.
Blue-detuned microwave dressing can additionally generate a repulsive shielding barrier that suppresses short-range collisional loss~\cite{Anderegg2021,Schindewolf2022,Bigagli2023}, while field-linked resonances provide further control of elastic and dipolar interactions~\cite{Chen2023}.
Accordingly, \(V_{ij}\) should be viewed as an engineered, Wannier-projected interaction kernel rather than the bare \(1/r^3\) dipole-dipole potential of statically polarized or tilted dipoles.

In such a lattice, the hopping \(t\) is set mainly by the lattice depth and spacing: a deeper lattice narrows the lowest band and reduces tunneling between neighboring Wannier orbitals.
The onsite interaction \(U\) is the coefficient of \(n_i(n_i-1)/2\), i.e., the extra energy for two molecules occupying the same Wannier orbital.
It is controlled by the onsite confinement, short-range molecular interactions, and the onsite matrix element of the dressed interaction; it should therefore be distinguished from the off-site tail \(V_{ij}\).
Finite onsite softness means that the \(n_i=2\) sector is allowed but energetically costly.
The off-site interaction \(V_{ij}\) is engineered by static-electric-field and microwave dressing of rotational states and projected onto Wannier orbitals on different lattice sites.
For the effective model studied here we use
\[
V_{ij}=\Vtwo\frac{x_{ij}^2-y_{ij}^2}{r_{ij}^5}+\Wsix\frac{1}{r_{ij}^6}.
\]
The first term is the sign-changing \(d_{x^2-y^2}\)-like channel that selects the axial \((q,0)\) stripe family, while the second term represents a short-range repulsive core that stabilizes the attractive axial channel.
Shielding provides the short-range stabilization and loss suppression represented by the repulsive core~\cite{Matsuda2020,Anderegg2021,Schindewolf2022,Bigagli2023}, while field-linked resonances offer an additional route for tuning elastic and dipolar interactions~\cite{Chen2023}.
Recent molecular condensation and droplet experiments show that coherent strongly interacting bosonic molecular fluids are now accessible~\cite{Bigagli2024,Zhang2026Droplets}.

The fixed-shape phase diagram is therefore a target path through one engineered interaction profile.
One first chooses a dressing geometry and shielding channel that set \(V_{ij}/\Vtwo\) and \(\Wsix/\Vtwo\); the lattice depth then changes \(\eta=\Vtwo/t\), with dressing or confinement co-tuned if needed to keep \(U/\Vtwo\) close to the desired value.
This is not an independent sweep of \(U\), \(\Vtwo\), and \(\Wsix\).
Ordinary contact-interacting alkali atoms would not realize this kernel, and Rydberg dressing would require additional angular engineering; the sign-changing off-site channel with a molecular repulsive core is the reason dressed polar molecules are the natural setting.
The hard-core counterpart selects the same axial stripe family~\cite{Zhang2026StripeFamily}; here we keep the onsite \(n_i\ge2\) sector and ask how it changes the doped stripe state.

At the central point used in the Letter, \(U/t=15\), \(\Vtwo/t=3.5\), and \(\Wsix/t=1.5\).
For representative lattice tunneling \(t/h\sim10\)--\(100\) Hz, this gives \(U/h\sim0.15\)--\(1.5\) kHz.
The off-site scales are then \(\Vtwo/h\sim35\)--\(350\) Hz and \(\Wsix/h\sim15\)--\(150\) Hz.
These are intended as order-of-magnitude targets, because the effective couplings depend on the molecular species, lattice spacing, dressing detuning, polarization geometry, and shielding channel.
The density can be varied by changing the total molecule number or the trap chemical potential.

In a smooth trap, the local chemical potential changes across the cloud, so the density-dependent trends shown in the Letter can be reconstructed from spatial regions with different local density.
This is the natural way to compare the hole side, the half-filled stripe parent, and the particle side in a single experimental setting.

The experimental phase diagram is reconstructed by combining standard cold-gas probes at the same local chemical potential~\cite{Bloch2008,Gross2021QuantumMicroscopy}.
In situ density profiles give \(n(\mu)\), plateaus, and local compressibility~\cite{Gemelke2009,Sherson2010}.
The selected stripe order is detected from the density structure factor, using Fourier analysis of snapshots~\cite{Sherson2010,Endres2011,Christakis2023MoleculeCorrelations}, coherent light scattering~\cite{Weitenberg2011}, or noise correlations~\cite{Folling2005,Rosenberg2022MolecularHBT}.
The relevant signature is a dominant \((q,0)\) peak, with transverse and diagonal channels used as checks.
The superfluid component is inferred from phase coherence and momentum-space interference, supplemented by spectroscopic probes of the superfluid--Mott transition~\cite{Greiner2002,Stoeferle2004}.
Thus HSS/LSS correspond to a plateau, a dominant \((q,0)\) stripe peak, and no coherent response, whereas DSS shows the same selected-stripe peak together with locally finite compressibility and phase coherence.

The soft-core defect channel can be probed by doublon-sensitive measurements. With site-occupation or parity-sensitive imaging, one can access local occupation statistics and density correlations in lattice gases~\cite{Sherson2010,Endres2011,Gross2021QuantumMicroscopy}; single-molecule imaging and correlation measurements have already been demonstrated in polar-molecule platforms~\cite{Rosenberg2022MolecularHBT,Christakis2023MoleculeCorrelations}.
In this language, the desired quantities are the local probabilities \(P_i(n)\), the integrated probability \(P_{\ge m}=L^{-2}\sum_i\sum_{n\ge m}P_i(n)\), and the onsite pair density \(D=L^{-2}\sum_i\avg{n_i(n_i-1)/2}\).

Double occupation can be probed through lattice-modulation spectroscopy or directly through number-resolved imaging that distinguishes singly and multiply occupied sites~\cite{Stoeferle2004,Hartke2020DoublonHole}.
The experimental prediction is not simply that double occupancy is present.
Rather, the particle-doped stripe state should show enhanced \(D\), finite phase coherence, and a persistent \((q,0)\) peak, while the commensurate hole-side LSS should retain stripe order but have strongly suppressed phase coherence and much smaller \(D\).
Low-momentum density fluctuations and real-space snapshots provide additional checks against macroscopic phase separation or local collapse.

\endgroup

\end{document}